\documentstyle[11pt,paspconf]{article}
\def\teff{{\sl T}_{\rm eff}}

\def\msol{{\sl M}_\odot}

\def\g{{\it g}\,}
\def\p{{\it p}\,}
\def\f{{\it f}\,}
\def\r{{\it r}\,}

\def\unity{ \hbox{1\kern-.23em l} }
\def\field{ \hbox{I\kern-.23em K} }

\begin{document}
\title{
       Any Recent Progress in the Theory of Pulsating Stars?
      }
\author{Alfred Gautschy}
\affil{Astronomisches Institut der Universit\"at Basel, Switzerland}

\begin{abstract}
To answer the topical question we survey synoptically the recent
literature in pulsation theory. We restrict the topics to those which
are not dealt with otherwise in this volume. We address research on
roAp stars, EC~14026 variables, strange modes, luminous blue variables,
pulsation-rotation coupling, and pulsations in compact objects seen
from the classical, as well as the relativistic, viewpoint.
\end{abstract}


\section{Introduction}
Well in accordance with the information era in which we 
live, variable-star research prospers, producing huge amounts
of high-quality data. For the various families of pulsators, a large
number of member stars are monitored regularly. Concerning theory,
this calls for new -- probably statistical -- avenues to be taken to
eventually improve our understanding of stellar physics and
evolution. We are finally no longer confined to detailed studies of
single objects only, but have access to statistically significant
ensembles. Therefore, conceptual advances are to be expected once the
ensemble aspect is given proper consideration in theoretical
studies. 

The last few years, however, have mostly seen the continuation of
traditional approaches to stellar pulsation theory. The contributions
from these approaches show that important basic aspects are still to
be clarified, and even discovered.  We have not yet reached a plateau
in our understanding with only minor quantitative bunny-hill problems
left to be solved.

The section 2 of this contribution deals with classical, Newtonian
stars for which formal tool-making aspects and excitation results are
discussed. Section 3 is devoted to compact objects for which we have
seen a considerable body of papers appearing recently. Much of the
excitement is generated by additional efforts in theory in connection
with the upcoming gravitational-wave detectors, and the recent launch
of the Chandra spacecraft.

\section{Newtonian Stars}

\subsection{Formal Developments}

Lee and Saio (1986, 1987) introduced a series expansion of the
latitudinal ($\theta$) dependence of pulsational perturbations in
rotating stars. The (usually rather low) number of terms in this
series representation always left some doubt about the quality of the
results, as it was not obvious if the dominant contributions were
included since the basis functions~--~associated Legendre
polynomials~--~generally do not converge quickly. This phase has
finally been overcome with the introduction of a direct integration
method for Laplace's tidal equation in the $\theta$ direction (Bildsten et
al. 1996, Lee \& Saio 1997).  Solving this additional eigenvalue
problem provides further sets of eigenvalues associated with modified
\g \,and \r \,modes and oscillatory convective modes, respectively.  An
important observation is that the \g \,modes' amplitudes become
increasingly more confined to low latitudes as the ratio of the
rotational to the oscillation frequency increases.  Oscillatory
convective modes, on the other hand, have only low amplitudes close to
the equator.  Lee \& Saio (1997) concluded positively that the
instability results for resonant couplings of envelope modes with
oscillatory convective modes in upper main-sequence stars (Lee \& Saio
1986) remain valid when using the new scheme, rather than the crude
two-term approximation used in the past.

For tidally disturbed stars in close binary systems, Savonije \&
Papaloizou (1997) went a step further, fully accounting for the
Coriolis force and casting the perturbation problem into a {\it
two-dimensional} boundary-value problem which was solved with a
complex-valued, finite-difference approach in both radial and latitudinal
directions.  The effect of the time-dependent external tidal potential
was included as a fixed-$\ell$ (quadrupole) perturbation with
adjustable orbital frequency. Based on this scheme, Witte \& Savonije
(1999) studied the tidal exchange of energy and angular momentum in a
rotating, massive, slightly evolved main-sequence star in 
an eccentric binary system. For various uniform rotation rates and series
of orbital periods, the resonances with \r \,and \g \,modes were
computed. These resonances appear to efficiently alter the orbital
evolution, as well as synchronize the massive star's rotation.

\subsection{Excitation of Pulsations}

The most celebrated pulsating variables of the recent past are clearly
the EC~14026 or variable sdB (sdBV) stars. Their popularity arose from
the rare incidence that the theoretical prediction appeared before the
publication of observational evidence of such objects (Charpinet et
al. 1996, Kilkenny et al. 1997). The observational data for this
class continues to grow rapidly (e.g. Koen et al. 1998a) and an instability
domain is beginning to emerge from these data. Theoretically, the instability
is clearly induced by the Z-bump in the opacity. The sdBVs have rather
large surface gravities and comparably high densities in their
envelopes. A consideration of the `opacity mountain', plotted over the
density-temperature plane, shows that the sharpness of the Z-bump
decreases as the density increases in the critical temperature
interval. Therefore, there is no further surprise to realize that
canonical heavy-element abundances are not sufficient to provide
enough driving to overcome radiative damping. Nevertheless, the
radiation field seems to be appropriate for radiative levitation to
induce a spatial distribution of heavier ions to sufficiently steepen
the Z-bump so that eventually the model stars become pulsationally
overstable in the observed temperature~--~$\log g$ domain
(e.g. Fontaine et al. 1998). Concerning the particularities of the
Z-bump, PG~1605+072 is of interest (Koen et al. 1998b). It shows about
a dozen, and possibly even more, oscillation modes with periods about in the
range of 200 to 540 s. These periods are a factor two to
three longer than what is found in the other sdBV stars.
Spectroscopic deduction of the stellar parameters indicates that
PG~1605+072 has a surface gravity of up to 0.5 dex lower than the rest
of the class. This might indeed hint at an enhanced excitation of \p \,
modes under lower-density conditions.

The success with the sdBV class fostered evolutionary and pulsational
studies of post-horizontal branch stars evolving towards the white
dwarf cooling tracks across the subdwarf domain. Therefore, the
prediction of gravity-mode instabilities induced by the
$\epsilon$-mechanism in the thin H-burning shell of low-mass stars
having settled close to the white dwarf cooling track is not
surprising (Charpinet et al. 1997).  Potential instabilities were
found for low-$\ell$ and low radial-order \g \,modes with periods
between 40 and 120 s. The instability region extends from about
4.64 to 4.88 in $\log \teff$.  The corresponding objects in the sky
would probably be classified as DAO stars. Currently, however,
there is no observational evidence for the existence of such variables.

Based on the coincidence of the thermal time-scale on top of a
prospective driving region and the pulsation period, it is clear that
the H~I/He~I ionization zone must play the dominant role in the
excitation of the high-order \p \,modes of roAp stars. Dziembowski \&
Goode (1996) also argued in this direction, but they were not
successful in actually finding overstable modes. After postulating a
chromosphere, Gautschy et al. (1998) could raise the outer edge of the
\p-mode cavity sufficiently high to eventually identify overstable
roAp-like acoustic modes. Homogeneous stellar envelopes allowed,
however, for rapidly oscillating Ap stars, as well as longer-period 
$\delta$ Scuti-like
modes to be excited.  Only after additionally introducing chemically
stratified envelopes (hypothetically caused by element sedimentation)
could they restrict the excited modes to the short-period domain. The
Gautschy et al. (1998) picture still has weak points, such as the 
absence of any observational evidence for chromospheres in roAp stars,
and too many
simultaneously excited modes of various spherical degrees.

The nature, and in particular the physics, of strange modes is still
debated. Their origin is becoming considerably clearer now. Buchler et al.
(1997) even found strange modes in Cepheid models. Their strange modes
can be analyzed even in the adiabatic limit, very much like in the
massive main-sequence star models of Glatzel \& Kiriakidis (1993) or
Saio et al. (1998). Finding strange modes in the adiabatic limit
removes much of their strangeness. It is seen that they owe their
existence to a sharp ridge in the acoustic cavity where waves with
appropriate oscillation frequency are effectively reflected, giving
birth to an additional oscillation spectrum of the stellar
envelope. Also, the large growth rates of these modes~--~let us call them
adiabatic strange modes~--~are easily explainable. They are confined
so much to the outermost stellar layers that the mode inertia is very
small. From quasi-adiabatic treatments we know that the imaginary part
of the eigenfrequency scales inversely with the mode
inertia. Therefore, even for a rather low driving efficiency, the
growth rate can become very large. These adiabatic strange modes are,
however, only part of the whole story. As pointed out in Saio et
al. (1998), there are stars with `nonadiabatic strange modes' that do
not show up in the adiabatic limit. In this latter case the cavity
splitting is believed to show up only in the fully nonadiabatic case.
In other words, for these modes, the interaction between the thermal
and the mechanical reservoir of the oscillator is essential. The
nonadiabatic strange modes are of particular interest, since they
contain all the `strange' mode physics, such as instability bands in
the nonadiabatic reversible limit. The crossings of adiabatic strange
modes with regular modes, on the other hand, unfold into avoided
crossings only. The detailed physics of the unfolding is still
unexplained.  A first step towards understanding the instability of
nonadiabatic strange modes was made by Saio et al. (1998). In a simple
model system, the radiation-pressure gradient is identified to drive
strange-mode instabilities very much like the gas-pressure gradient does
for dynamical instabilities. Numerical experience shows indeed that
nonadiabatic strange modes pop up whenever the model stars show
strongly radiation-dominated layers.

Partly in connection with strange modes in massive stars a 
well-documented controversy between Glatzel \& Kiriakidis
(1998) and Stothers (Stothers \& Chin 1993, Stothers 1999) developed
around the concept of dynamical instability. Stothers (1999)
attributed nonlinear oscillatory instabilities of massive model-star
envelopes to the nonadiabatic manifestation of an adiabatically
diagnosed $\langle \Gamma \rangle < 4/3$ instability. The brackets
denote a suitably defined spatial average; this suitability was also a
point of controversy.  Glatzel \& Kiriakidis (1993) criticized
dynamical-instability claims by Stothers \& Chin (1993) and argued for
the necessity of a fully nonadiabatic treatment to understand LBV
variability and eruptions in these stars.  Even if the nonadiabatic,
nonlinear simulations of dynamically unstable models (computed with
the adiabaticity constraint) are strongly pulsationally unstable, it
is formally unclear how to connect the pulsational with the
dynamical instability. In a system in which the time-scale of energy
exchange and the time-scale for sound propagation through the system
become comparable, the concept of dynamical instability loses its
foundation.  Adopting, however, a more pragmatic point of view, we can
certainly say that whenever we come across a stellar model with a
dynamically unstable adiabatic fundamental mode, it is worth a closer
look as it is obviously hardly bound anymore, even in a
thermodynamically more appropriate framework.
 
Based on the Chandrasekhar--Milne expansion of the mechanical
structure equations for a rotating star, Lee (1998) investigated the
influence of rotational deformation on the stability of axisymmetric
acoustic and gravity modes in B-type main sequence stars.  Acoustic
nonradial modes were found to be stabilized at high rotation
frequencies, i.e. for rotation speeds approaching break-up
speed. Quasi-radial modes, however, remained overstable even in the
very rapid rotators. The results were, therefore, comparable with the
conclusions by Lee \& Baraffe (1995), who studied the same phenomenon
for non-axisymmetric perturbations. Lee's (1998) explorative
nonadiabatic investigation of low-frequency \g \,modes, representative
for the pulsation modes in slowly pulsating B stars (SPBs), led him to
conclude that they are unlikely to be damped out by the star's
rotation, despite the fact that the importance of rotation, measured
by the ratio of the rotational to the oscillation frequency, is higher
for the \g \,modes than for the acoustic modes. Quasi-adiabatic
eigensolutions by Ushomirsky \& Bildsten (1998) showed, on the other
hand, that \g \,modes appropriate for SPBs could be damped by
rotation. Furthermore, oscillation modes which were found to be stable
in non-rotating models could be spun up to overstability.  Ushomirsky
\& Bildsten (1998) argued that it is the period measured in the
co-rotating frame of reference which is the important quantity to fit
the thermal time-scale of the envelope above the driving region. Lee's
(1998) view is different; he argues that the mean radius of the
equipotential surfaces, introduced in his analyses, shifts the
excitation region to larger effective radii as the rotation speed of a
star increases. Thereby, the effective period which can match the
thermal time-scale envelope overlying the excitation region
drops. This effect was found to be more pronounced for the \p \,modes
in $\beta$ Cephei stars than for the \g \,modes in SPBs. In contrast
to the Ushomirsky \& Bildsten (1998) analysis, which relies on the
direct solution of the tidal equation, Lee (1998) used a two-term
expansion in the $\theta$-dependence of the perturbations which might
not be fully adequate for the problem.  In other words, the
quasi-adiabatic analyses versus the Legendre-polynomial expansions in
latitude have not yet converged to a common prediction.  In any case,
it appears to be clear that rotating models are necessary to fully
understand the observed populations of blue main-sequence pulsators,
particularly in stellar clusters. It might well be that rotation
removes some of the instabilities found in non-rotating models. Rapid
rotators might also be unstable at lower effective temperatures than
non-rotating stars. As stellar rotation constricts the amplitudes of
\g \,modes to lower stellar latitudes, the lack of rapid rotators
observed among SPB stars (e.g. Balona \& Koen 1994) might possibly be
attributable to an observational selection effect in the end.

The behavior of \r \,modes in differentially rotating stellar envelopes
was studied by Wolff (1998) in connection with solar oscillations. The
geostrophic mode with vanishing oscillation frequency measured in the
co-rotating frame picks up a finite oscillation frequency in the
differential rotation case. The \r-mode spectrum consists then of a
slow and a fast branch. Depending on the magnitude of rotational shear
and the generalized spherical degrees of rapid and slow \r \,modes, their
characters can converge. They even can merge in frequency space (as a
function of shear parameterization) and thereafter cease to exist.

\section{Compact Objects}

As white dwarfs cool, the central density eventually rises
sufficiently for the stellar matter to pass through a first-order
phase transition to crystallize. The existence of a crystalline
interior has consequences for the interpretation of the white dwarfs'
luminosity function and therefore their dating.  As DA white dwarfs
can show up as \g mode pulsators whose instability region~--~measured
in the $\rho$-$T$ plane~--~extends into the domain where
crystallization could have started, they are prospective candidates to
search for observable consequences of a solid interior. Indeed, the
DAV star BPM~37093 is the prime candidate for such
investigations. Bradley (1996) contemplated that crystallization
should manifest itself in the rate of period change which differs from
a star with a Coulomb-fluid interior, first because of the release of
latent heat at crystallization and later on due to Debye cooling. The
monitoring time necessary to detect such an effect would be very long
-- of the order of ten to twenty years. The effect dominating the
period change might, however, be the growing of the crystallized
sphere in the star's interior (Winget et al. 1997). The latter authors
suggested, based on a simplified treatment of the linear eigenvalue
problem, the measurement of the period spacings of identified \g
\,modes in stars like BPM~37093 to deduce the magnitude of
crystallization. They found that the mean period spacing, measured
relative to an uncrystallized star with otherwise identical
properties, increases by up to about 30\%, depending on the mass
fraction of the crystallized interior. As the mean period spacing of
\g \,modes also depends on other stellar parameters, such as thickness
of H and He layers and position on the HR plane, complementary mode
properties have to be found to measure crystallization conclusively.

A more favorable environment to observe effects of solid stellar
matter was discussed by Duncan (1998). He suggested a search for
seismic toroidal modes excited during star-quakes in neutron stars.
If soft gamma repeaters (SGR) are highly magnetized neutron stars
experiencing fracture events in their solid crusts, global seismic
oscillations can be excited very much like on Earth.  As the crust is
rather superficial, seismic modes might attain significant amplitudes
on the surface to make them observable. Duncan (1998) therefore
proposed a search for periodicities in SGR signals from some ten Hz
upwards.

Considerable efforts have gone into studying relativistic oscillations of
compact objects. A huge body of data exists showing time-variable
phenomena in the frequency domain from Hz to kHz which are, in principle, 
attributable to oscillations on neutron stars. Much of the present
enthusiasm originated from the prospect to observe such signatures in
the forthcoming gravitational wave detectors (Owen et al. 1998).

The Chandrasekhar--Friedman--Schutz (CFS) instability is known to
destabilize spheroidal modes in inviscid rotating relativistic
bodies. Under the influence of viscous forces, only \f \,modes with
$\ell = m = 2$ survive in objects rotating close to break-up speed
(e.g. Lindblom 1995). Therefore, the CFS instability was considered to be of 
little importance in nature. This situation changed when Andersson
(1998) found that toroidal modes in rotating stars are strongly
overstable to the CFS gravitational-radiation reaction.  The
instability grows proportional to $(\Omega \sqrt{R^3/M})^{10}$ for
sectoral dipole modes; $\Omega$ stands for the rotation rate of a
star with mass $M$ and radius $R$.  Friedman \& Morsink (1998) proved
that Andersson's numerical finding applies to any relativistic
rotating body.  For most of the \r \,modes, the CFS instability persists
at any value of $\Omega$. Hence, the severe constraint of very rapidly
rotating relativistic bodies associated with spheroidal modes
disappeared. Nevertheless, to estimate which modes could survive in
realistic neutron stars, the effects of rather uncertain viscosity
effects needed to be studied. First attempts revealed that \r-modes might
survive in the temperature window between $10^9$ and $10^{10}$~K.
Below the lower boundary, shear viscosity damps the \r-modes, and above
about $10^{10}$~K bulk viscosity over-compensates the CFS
instability. In this scenario, young hot, rapidly spinning neutron
stars are expected to pass through the instability window within the
first few years after their birth, losing up to $0.01 \msol c^2$ of
energy, thereby slowing their rotation rate to five to ten percent
of the break-up rotation speed.  This mechanism could explain why some
young pulsars are observed to rotate slowly compared with what
is expected from angular-momentum conservation during
collapse. Furthermore, the \r-mode instability would not permit
milli-second pulsars to form via accretion-induced collapse of white
dwarfs. The white dwarfs would get too hot in the process and lose
much of their angular momentum through the CFS instability. Instead,
milli-second pulsars must be formed by means of accretion onto a
neutron star, keeping the recipient object always at sufficiently low
temperature to suppress the \r-mode instability.  Observationally,
gravitational wave radiation generated by the CFS instability is now
expected to be detectable by enhanced LIGO interferometers to
distances as far as  Virgo cluster (Owen et al. 1998).

The \r-modes referred to in the last paragraph are only one sub-class
of inertial modes that owes their existence to rotation. Others, for
which the spheroidal contribution to the eigensolutions is of
comparable strength as the toroidal one were investigated for CFS
instability by Yoshida \& Lee (1999). They find that the `inertial
modes' in these stars are also CFS unstable, but less so than the
\r-modes.  Nevertheless, the most overstable inertial modes appear to
survive even in the dissipative case.

Thermally or compositionally stratified neutron stars also possess
\g \,modes with sufficiently long periods to satisfy the
condition for CFS instability. Lai (1999) investigated this case and
concluded that \g \,modes become overstable by means of the CFS
instability if the rotation rate is comparable or larger than the
considered \g-mode frequency.  Viscosity seems to extinguish the
instability, except possibly around temperatures of $10^9$~K.  The
\g-mode instability is, in any case, orders of magnitude weaker than the
inertial-mode instability and might, therefore,
 not be of importance in nature.

Gravity-mode oscillations in neutron stars have been a focus of
numerous studies in the past (see e.g. Gautschy \& Saio 1995). Lately,
Bildsten \& Cumming (1998) added a new prospective \g-mode cavity to
the theoretical picture. They proposed an additional class of \g \,modes
in matter-accreting neutron stars. These modes owe their existence to
the compositional inhomogeneity developing at the base of the hydrogen
and helium layer that builds up on the surface and transmutes into
iron-group elements by unstable H/He burning and by electron captures.
Very much like at the outer edge of the convective cores of massive
main-sequence stars, a sharp gradient in the molecular weight builds
up which acts as a cavity for the previously mentioned \g \,modes and an
interface mode. In the case of the upper main-sequence stars this kind
of gravity mode was baptized core \g \,modes. In the case of neutron
stars the cavity lies rather close to the surface, actually at the
base of the superficial suprafluid ocean. We refer to modes being
essentially trapped in this composition interface as interface \g
\,modes.  In addition, if the neutron star's background is not 
isentropic, the non-vanishing Brunt--V\"ais\"al\"a frequency permits the
`normal' thermal buoyancy \g modes to show up. The resulting frequency
spectrum can then become rather intricate as the frequency domains of
the interface and thermal \g modes can overlap. In the adiabatic mode
treatment of Bildsten \& Cumming (1998) the resulting mode interaction
between the two families unfolds into avoided crossings.

The complicated nuclear physics and thermodynamics in neutron stars
can produce a multitude of narrow \g \,mode cavities by density jumps
alone. Much of the attention given to the interface \g \,modes at the
bottom of the superficial suprafluid originates from the frequencies
of these modes which are compatible with observed QPOs in the
frequency domain of a few times 10 Hz. The adiabatic properties
of the interface modes in slowly rotating neutron stars have been well
discussed. The case of rapid rotation~--~measured relative to the \g
\,mode frequency~--~has still to be tackled in more detail. In
particular, however, the excitation mechanism of the interface \g
\,modes and \f-type interface mode have to be addressed. As the
interface modes are well trapped in the compositional transition
region, they will hardly see anything of the overlying H/He burning.
A conceivable driving agent could, however, be an electron-capture
instability of the transition layer becoming oscillatory by
nonadiabaticity so close to the neutron star's surface.

\acknowledgments The Swiss National Science Foundation supported the
author via a PROFIL2 fellowship during the writing of this review and
Vienna's Caf\'e Eiles provided a comfortable Bohemian atmosphere to
finalize it.  As usual, H. Harzenmoser gave selflessly
away his knowledge of the literature collected in his unequaled
brain-oriented database. Finally I would like to thank L. Szabados
and D. Kurtz for their editing efforts.


\begin{references}
\reference Andersson, N. 1998, 
	ApJ 502, 708
\reference Balona, L., \& Koen, C. 1994,
	MNRAS 267, 1071
\reference Bildsten, L., \& Cumming, A. 1998,
	ApJ 506, 842
\reference Bildsten, L., Ushomirsky, G., \& Cutler C. 1996,
	ApJ 460, 827
\reference Bradley, P. A. 1996,
	ApJ 468, 350
\reference Buchler, J. R., Yecko, P. A., \& Koll\'ath, Z. 1997,
	A\&A 326, 669
\reference Charpinet, S., Fontaine, G., Brassard, P., \& Dorman, B. 1996, 
        ApJ 471, L103
\reference Charpinet, S., Fontaine, G., \& Brassard, P. 1997, 
        ApJ 489, L149
\reference Duncan, R. C. 1998, 
	ApJ 498, L45
\reference Dziembowski, W. A., \& Goode, P. R. 1996, 
	ApJ 458, 338
\reference Fontaine, G., Charpinet, S., Brassard, P., Chayer, P., 
        Rogers, F. J., et al. 1998, in
	New Eyes to See Inside the Sun and the Stars, 
        IAU Symp. 185, 
	Eds. F.-L. Deubner et al., Dordrecht, p. 367
\reference Friedman J. L., \& Morsink S. M. 1998,
	ApJ 502, 714
\reference Gautschy, A., \& Saio, H. 1995,
	ARAA 33, 75
\reference Gautschy, A., Saio, H., \& Harzenmoser, H. 1998,
        MNRAS 301, 31
\reference Glatzel, W., \& Kiriakidis, M. 1993, 
	MNRAS 262, 85
\reference Glatzel, W., \& Kiriakidis, M. 1998,
 	MNRAS 295, 251
\reference Kilkenny, D., Koen, C., O'Donoghue, D., \& Stobie, R. S. 1997, 
  	MNRAS 285, 640
\reference Koen, C., O'Donoghue, D., Kilkenny, D., \& Stobie, R. S. 1998a,
        in 
	New Eyes to See Inside the Sun and the Stars, 
        IAU Symp. 185, 
	Eds. F.-L. Deubner et al., Dordrecht, p. 361
\reference Koen, C., O'Donoghue, D., Kilkenny, D., Lynas-Gray, A. E., 
        Marang, F., \& van Wyk, F. 1998b
        MNRAS 296, 317
\reference Lai, D. 1999,
	MNRAS in press
\reference Lee, U. 1998, 
	ApJ 497, 912
\reference Lee, U., \& Baraffe, I. 1995, 
	A\&A 301, 419
\reference Lee, U., \& Saio, H. 1986, 
	MNRAS 221, 365
\reference Lee, U., \& Saio, H. 1987, 
	MNRAS 224, 513
\reference Lee, U., \& Saio, H. 1997, 
	ApJ 491, 839
\reference Lindblom, L. 1995,
	ApJ 438, 265
\reference Owen, B. J., Lindblom, L., Cutler, C., Schutz, B. F., 
        Vecchio, A., \& Andersson, N. 1998, 
	Phys. Rev. D, 58, 084020
\reference Saio, H., Baker, N. H., \& Gautschy, A. 1998, 
        MNRAS 294, 622, 
\reference Savonije, G. J., \& Papaloizou, J. C. B. 1997,
        MNRAS 291, 663
\reference Stothers, R. B. 1999, 
        MNRAS 305, 365
\reference Stothers, R. B., \& Chin, C. W. 1993,
	ApJ 408, L85
\reference Ushomirsky, G., \& Bildsten, L. 1998,
	ApJ 497, L101
\reference Yoshida, S., \& Lee, U. 1999,
	submitted to ApJ
\reference Winget, D. E., Kepler, S. O., Kanaan, A., 
	Montgomery, M. H., \& Giovannini, O. 1997,
	ApJ 487, L191
\reference Witte, M. G., \& Savonije, G. J. 1999, 
	A\&A 341, 842
\reference Wolff, C. L. 1998, 
        ApJ 502, 961
\end{references}
\end{document}